\documentclass[aps,prx,amsmath,amssymb,reprint,longbibliography,nofootinbib]{revtex4-1}
\usepackage{ifthen}
\usepackage{cancel}
\usepackage{epstopdf}
\newboolean{acs}\setboolean{acs}{false}
\newcommand{\chooseformat}[2]
{
	\ifthenelse{\boolean{acs}}
	{#1}{#2}
} 

\usepackage[uline]{hhtensor}
\usepackage{overpic}
\usepackage{booktabs}
\usepackage{graphicx}

\usepackage{bm}

\usepackage[caption=false]{subfig}

\newcommand\HH{\mathbf{H}}
\newcommand\EE{\mathbf{E}}
\newcommand\p{\mathbf{p}}
\newcommand\rrp{\mathbf{r_0}}
\newcommand\pbar{\mathbf{\bar{p}}}
\newcommand\ed{\mathbf{d}}
\newcommand\md{\mathbf{m}}
\newcommand\Fp{F_\p^\pbar{\tiny(+)}}
\newcommand\Fm{F_\p^\pbar{\tiny(-)}}
\newcommand\MdE{{\matr{\alpha}}_{\ed\mathbf{E}}}
\newcommand\MdH{{\matr{\alpha}}_{\ed\mathbf{H}}}
\newcommand\MmE{{\matr{\alpha}}_{\md\mathbf{E}}}
\newcommand\MmH{{\matr{\alpha}}_{\md\mathbf{H}}}

\newcommand\ddd{d_{\p}^{\pbar}}
\newcommand\aaa{a_{\p}^{\pbar}}
\newcommand\bbb{b_{\p}^{\pbar}}
\newcommand\ccc{c_{\p}^{\pbar}}

\newcommand\zhat{\mathbf{\hat{z}}}

\newcommand\Deltabeta{\Delta\beta_\p^\pbar}
\newcommand\Deltaeta{\Delta\eta_\p^\pbar}
\newcommand\betahat{\hat{\beta}_\p^\pbar}
\newcommand\etahat{\hat{\eta}_\p^\pbar}
\newcommand\psippbar{\psi_\p^\pbar}

\newcommand\shortminus{\scalebox{0.75}[1.0]{\( - \)}}
\RequirePackage[normalem]{ulem} \RequirePackage{color}\definecolor{RED}{rgb}{1,0,0}\definecolor{BLUE}{rgb}{0,0,1}                                  
\chooseformat{
\title{Dual and chiral objects for optical activity in general scattering directions}\author{Ivan Fernandez-Corbaton}\email{ivan.fernandez-corbaton@kit.edu}
\affiliation{Institute of Nanotechnology, Karlsruhe Institute of Technology, 76021 Karlsruhe, Germany}
\author{Martin Fruhnert}\affiliation{Institute of Condensed Matter Theory and Solid State Optics, Friedrich-Schiller-Universit\"at, Max-Wien-Platz 1, 07743 Jena, Germany}
\author{Carsten Rockstuhl}\affiliation{Institut f\"ur Theoretische Festk\"orperphysik, Karlsruhe Institute of Technology, 76131 Karlsruhe, Germany}
\alsoaffiliation{Institut of Nanotechnology, Karlsruhe Institute of Technology, 76021 Karlsruhe, Germany}
\begin{document}
}
{\begin{document}
\title{Dual and chiral objects for optical activity in general scattering directions}\author{Ivan Fernandez-Corbaton}\email{ivan.fernandez-corbaton@kit.edu}\affiliation{Institut of Nanotechnology, Karlsruhe Institute of Technology, 76021 Karlsruhe, Germany}
\author{Martin Fruhnert}\affiliation{Institute of Condensed Matter Theory and Solid State Optics, Friedrich-Schiller-Universit\"at, Max-Wien-Platz 1, 07743 Jena, Germany}
\author{Carsten Rockstuhl}\affiliation{Institut f\"ur Theoretische Festk\"orperphysik, Karlsruhe Institute of Technology, 76131 Karlsruhe, Germany}
\affiliation{Institut of Nanotechnology, Karlsruhe Institute of Technology, 76021 Karlsruhe, Germany}}

\begin{abstract}
	Optically active artificial structures have attracted tremendous research attention. Such structures must meet two requirements: Lack of spatial inversion symmetries and, a condition usually not explicitly considered, the structure shall preserve the helicity of light, which implies that there must be a vanishing coupling between the states of opposite polarization handedness among incident and scattered plane waves. Here, we put forward and demonstrate that a unit cell made from chiraly arranged electromagnetically dual scatterers serves exactly this purpose. We prove this by demonstrating optical activity of such unit cell in general scattering directions.
\end{abstract}
\maketitle

Research on optical activity started with the works of Arago \cite{Arago1811}, Biot \cite{Biot1815}, and Pasteur \cite{Pasteur1848}, who studied the rotation of the polarization of light upon propagation through some crystals and molecular solutions. Pasteur identified the absence of mirror planes of symmetry of the molecule as a necessary condition for optical activity. Optical activity is nowadays a vast field of fundamental and applied research across physics, chemistry and biology \cite{Barron2004,Guerrero2011,Hentschel2012,Frank2013,Meinzer2013}.

The ability of some natural systems to rotate the polarization of light has been artificially reproduced in two dimensional planar arrays of strongly scattering unit cells \cite{Zhang2009,Plum2009,Decker2010}. There, the polarization of a normally incident field is rotated in transmission, the forward scattering direction of the system. To observe this effect, the array must lack reflection symmetry across all planes perpendicular to it. Research in artificial optical activity in non-forward scattering directions has shown that, even though the systems also break the necessary spatial inversion symmetries, the resulting transformation of the polarization is qualitatively different from the one obtained in the forward direction \cite{Papakostas2003,Ren2012}. The difference is that, in the forward direction, a linearly polarized field has its polarization rotated by a constant angle, independent of the incident polarization angle. In non-forward scattering, however, the amount of rotation depends explicitly on the angle of the incident linear polarization (\chooseformat{\onlinecite{Fig. 2}{Papakostas2003}}{\cite[Fig. 2]{Papakostas2003}},\chooseformat{\onlinecite{Fig. 5}{Ren2012}}{\cite[Fig. 5]{Ren2012}}). Such kind of polarization transformation {\em does not meet} the definition of optical activity in terms of circular birefringence \chooseformat{\onlinecite{Sec. 1}{Condon1937}, \onlinecite{Chap. 1.2}{Barron2004}}{\cite[Sec. 1]{Condon1937},\cite[Chap. 1.2]{Barron2004}}. This definition includes the possibility of different absorption of the two circular polarization handedness (circular dichroism), and therefore allows the output polarization to become elliptical. Nonetheless, this ellipse must rotate in a consistent manner as the incident polarization angle changes: The output ellipticity and the relative rotation angle of its major axis {\em shall be independent} of the incident polarization angle. 

The explanation for the observed qualitative difference between the forward and non-forward directions is that, contrary to what is often stated \cite{Oloane1980,Barron2004,Bishop1993}, breaking spatial inversion symmetries is {\em not the only necessary condition} for optical activity \cite{FerCor2012c}. Besides breaking spatial inversion symmetries, optical activity also requires, as an additional condition, that polarization handedness be preserved in the scattering process. This means that the coupling between incident and scattered components of different polarization handedness must be zero. This requirement is in addition to the lack of mirror symmetry across the scattering plane, and is the necessary and sufficient condition for the output rotation angle and ellipticity to be independent of the incident linear polarization angle (see Ref. \cite{FerCor2012c}, and the \chooseformat{Supp. Info.}{Appendix \ref{app:optact}} of this article). 

The coupling between different polarization handedness can be discussed within the framework of symmetries and conservation laws by means of the helicity of the field \cite{FerCor2012p,FerCorTHESIS}. For a plane wave, helicity can be defined as the polarization handedness with respect to its momentum vector. The forward scattering direction is special in the sense that helicity preservation can be achieved by purely geometrical means with scatterers possessing discrete rotational symmetry with degree higher than two \cite{Hu1987,Menzel2010,FerCor2013c,Kaschke2014}. In particular, the disorder induced effective cylindrical symmetry of a solution of randomly oriented chiral molecules ensures helicity preservation and, together with the inherent chirality of the molecules, allows optical activity in the forward direction. This geometrical helicity preservation is achieved in the forward scattering of planar arrays with four-fold \cite{Zhang2009,Plum2009,Decker2010,Kaschke2012} and three-fold \cite{Kaschke2014} rotational symmetry.

The arguments that lead to this geometrical helicity preservation involve the angular momentum of the plane waves and rely on the fact that the incident and scattered directions share the same axis. These arguments {\em do not apply} to a general scattering direction \cite{FerCor2013c}, and the components of different polarization handedness will usually mix in non-forward scattering. This explains why a chiral system does not generally exhibit optical activity in non-forward directions: It does not generally meet the second necessary condition. However, there is at least one way to achieve helicity preservation also in general non-forward directions, and, as far as we know, it is the only way. In the same fundamental sense in which rotational symmetry of the scatterer ensures the preservation of angular momentum, electromagnetic duality symmetry of the scatterer ensures helicity preservation in all directions.

Chiral and dual objects break all mirror planes of symmetry (due to chirality) and preserve helicity for all incident/scattered plane waves (due to duality symmetry). They are hence appropriate for achieving optical activity in general scattering directions. In general, a chiral and dual object will exhibit different amounts of polarization rotation for different pairs of incident/scattered directions, while meeting the definition of optical activity for each pair. This is consistent with the fact that, as far as we know, there is no reason to expect a constant rotation angle in the general case.

\section{Outline}
In this article, we incorporate the requirement of helicity preservation into the design of artificial optical activity in general scattering directions and address it through the duality symmetry properties of the scatterer. To such end we employ appropriately chosen small dielectric spheres. While their materials are not dual symmetric according to the macroscopic Maxwell's equations \cite{FerCor2012p}, the electromagnetically small spheres are dual symmetric in the dipolar approximation for a carefully chosen set of parameters at a given design wavelength \cite{Zambrana2013b}, and allow us to meet the design requirement up to good approximation. We show that a wavelength sized chiral structure composed of four different dipolarly dual spheres exhibits optical activity in general scattering directions. Motivated by what is feasible with self-assembly nanofabrication technologies \cite{Boal2000,Fan2011,Malassis2013}, we have chosen for the chiral structure a tetrahedral arrangement \cite{Mastroianni2009}. Other chiral compositions can be considered as well, like for example helical arrangements of nanoparticles which can also be fabricated with self-assembly techniques \cite{Kuzyk2012,Schreiber2013}. The dipolarly dual spheres are dielectric particles with high refractive index. Such particles are increasingly being considered as building blocks for optical antennas, metamaterials, and, in general, field manipulation devices \cite{Pellegrini2009,Krasnok2012,Filonov2012,Schuller2007}. This is promoted by their negligible absorption and their ability to sustain Mie-type resonances, enabling strong light-matter interaction and, in particular, a notable electric and magnetic dipolar response \cite{Person2013,Fu2013,Schuller2007,Schmidt2014}.

We compare the optical activity properties of the chosen structure to those of two other chiral tetrahedral structures: One made with (ideal) dual symmetric materials and another one made with spheres whose parameters set them away from the condition of dipolar duality. The comparison illustrates the importance of duality symmetry in optical activity. 

The article is structured as follows. In the next section we explain the design in detail and specify the candidate structure. We also specify two other structures which are used for comparison purposes. In the following section, we outline a general methodology for the analysis of the optical activity properties of a scatterer. We apply the methodology to the three structures and discuss the results. We finish with the conclusion. In this article, the electromagnetic responses of the structures are calculated through rigorous techniques which allow to compute their scattering matrices to an arbitrary multipolar order \cite{Mackowski1996,Xu1995,Muhlig2010,Fruhnert2014}. We use order nine, which provides sufficient convergence, and renders our analysis and conclusions valid up to the approximations inherent in the macroscopic Maxwell's equations.

\begin{figure}[h!]
        \centering
		\captionsetup[subfigure]{labelformat=empty}
	\subfloat[]{\label{fig:scatt}
	\chooseformat{\includegraphics[width=4.5cm]{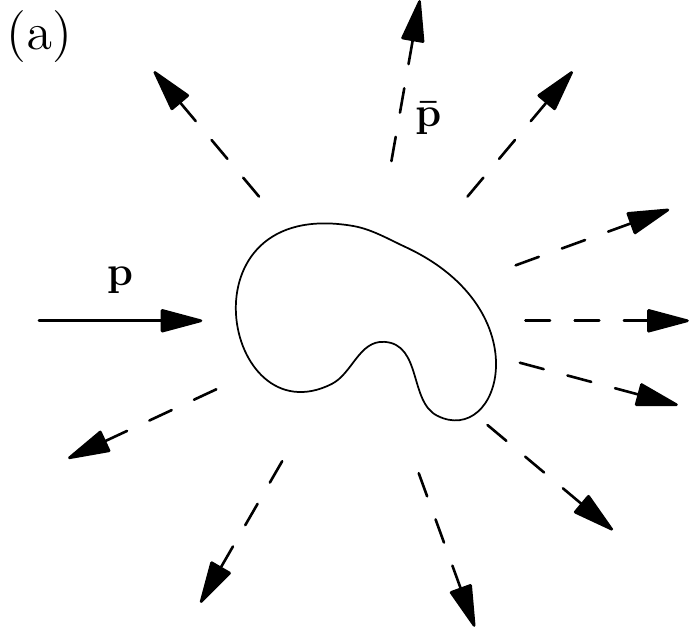}}{\includegraphics[width=4.5cm]{scatt}}}\\
		\subfloat[]{\label{fig:poltrans} 
		\chooseformat{\includegraphics[width=8cm]{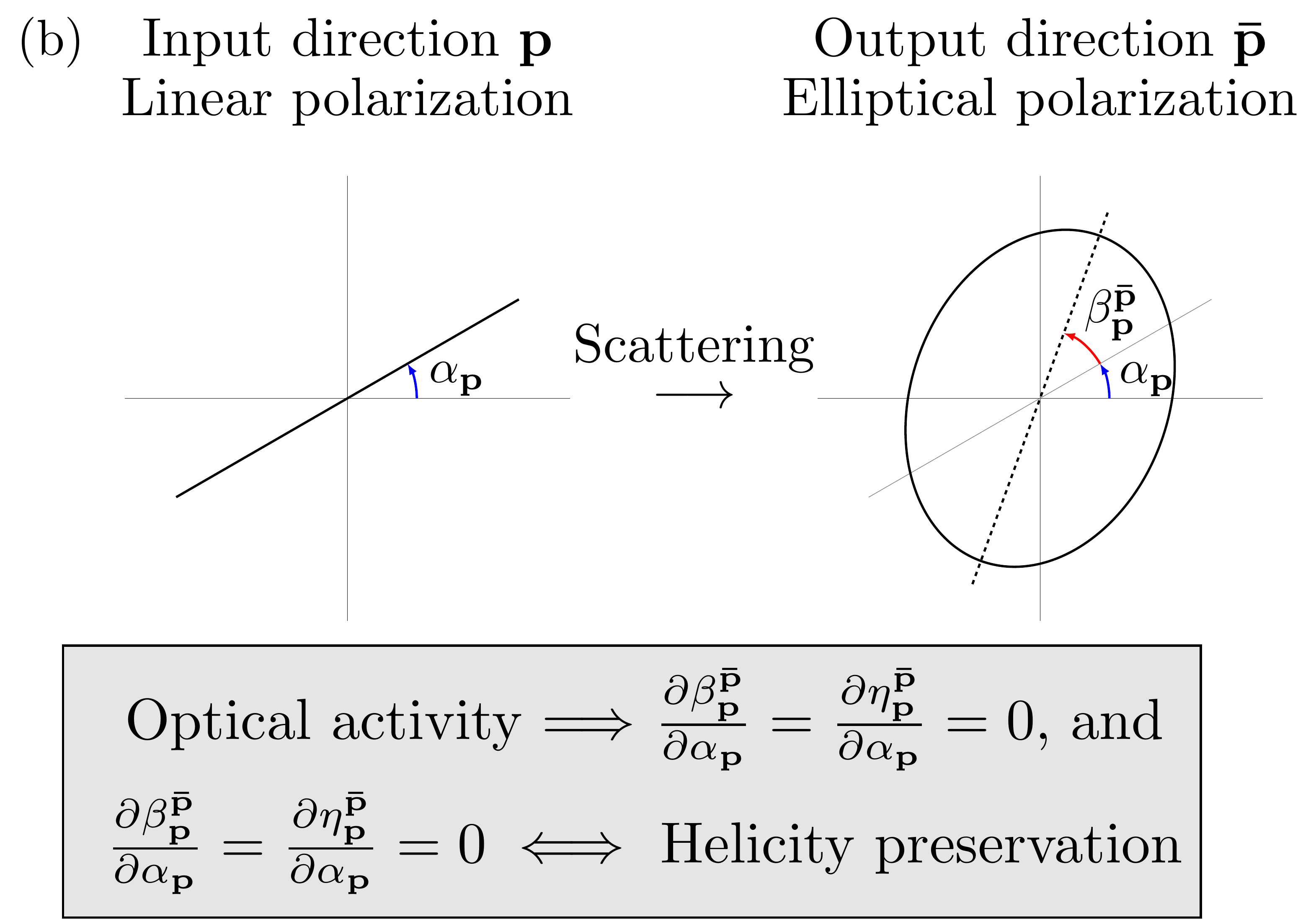}}{\includegraphics[width=8cm]{pol_rot_p_pbar}}}

		\caption{\label{fig:setting} Panel (a): An object scatters an incident plane wave (continuous arrow) into many scattered plane waves (dashed arrows). Optical activity for two arbitrary scattering directions like $\p/\pbar$ implies a polarization transformation of the kind illustrated in panel (b). Panel (b): Upon scattering, the incident linear polarization turns into elliptical polarization with ellipticity $\eta_\p^\pbar$ and main axis rotated by an angle $\beta_\p^\pbar$ with respect to the input polarization angle $\alpha_\p$. As $\alpha_\p$ varies, both $\beta_\p^\pbar$ and $\eta_\p^\pbar$ stay constant. This constant behavior is only achieved when the scatterer does not couple states of different polarization handedness (helicity). This necessary condition for optical activity is in addition to the required lack of mirror reflection symmetry of the scatterer across the plane(s) containing $\p$ and $\pbar$. The preservation of helicity can be achieved by geometrical means if $\p=\pbar$ (forward scattering). For the general case, electromagnetic duality symmetry of the scatterer ensures that helicity is preserved for all $\p/\pbar$.}
		
\end{figure}

\section{Design of optical activity in general scattering directions}\label{sec:des}
We set out to design a structure exhibiting optical activity for general pairs of incident and scattered directions, e.g. in non-forward scattering. Let us consider an arbitrary pair of incident/scattered directions labeled by the momentum vector of their corresponding plane waves $\p$/$\pbar$ (see Fig. \ref{fig:setting}). A necessary condition for the polarization of $\pbar$ to be a rotated version of that of $\p$ (in the sense of Fig. \ref{fig:poltrans}), is that the scatterer lacks mirror reflection symmetry across the scattering plane, that is, across the plane defined by the two vectors $\p$ and $\pbar$. A short proof of this intuitive result can be found in Ref. \cite{FerCor2012c}. If the two momentum vectors are parallel ($\p=\pbar$, forward scattering), the scatterer must lack reflection symmetry across all the planes containing them, as is the case for the arrays in Refs. \cite{Zhang2009,Plum2009,Decker2010}. Since we want to achieve optical activity in general scattering directions, we choose the structure to be chiral, ensuring the breaking of all mirror planes of symmetry. The additional requirement of helicity preservation is addressed through the duality symmetry properties of the scatterer. A scatterer is said to be dual symmetric if its electromagnetic response is invariant under the electromagnetic duality transformation. Duality acts on the electric ($\EE$) and magnetic ($\HH$) fields \chooseformat{\onlinecite{Eq. 6.151}{Jackson1998}}{\cite[Eq. 6.151]{Jackson1998}}:

\begin{equation}
	\label{eq:duality}
\begin{split}
\EE\rightarrow \EE_\theta&=\EE\cos\theta  - Z\HH\sin\theta , \\
Z\HH\rightarrow Z\HH_\theta&=\EE\sin\theta + Z\HH\cos\theta,
\end{split}
\end{equation}
where $\theta$ is an arbitrary real angle and $Z$ a reference impedance.

Dual symmetric scatterers preserve helicity for all $\p$/$\pbar$. \chooseformat{The Supp. Info.}{Appendix \ref{app:heldual}} contains a brief introduction to helicity and duality. Their use in light matter interaction problems is discussed in detail in \cite{FerCorTHESIS}. 

Non-dual symmetric scatterers change the helicity of the field interacting with them. We now introduce a measure of helicity change (duality breaking) for an arbitrary object. To this end, we consider the scattering matrix of the object expressed in a basis of electromagnetic modes with helicity as the polarization index ($\lambda=\pm1$). The other labels needed to identify each basis mode are lumped into a collective index\footnote{The contents of $\eta$ depend on the further choice of basis. For example, $\eta$ contains the three components of momentum for plane waves of well defined helicity, or, the frequency, total angular momentum and angular momentum for multipolar fields of well defined helicity.} $\eta$. The scattering coefficient $s_{\eta,\lambda}^{\bar{\eta},\bar{\lambda}}$ is then the component of the scattered field in the $(\bar{\eta},\bar{\lambda})$ mode resulting from the interaction of the object with an incident $(\eta,\lambda)$ mode. We define the relative helicity change $\cancel{D}$ as the ratio between the sum of the modulus square of all the helicity flipping scattering coefficients and the sum of the modulus square of all the scattering coefficients: 

\begin{equation}
	\label{eq:cancelD}
	\cancel{D}=\frac{\sum_{\eta}\sum_{\bar{\eta}}\sum_{\lambda=\pm 1}|s_{\eta,\lambda}^{\bar{\eta},-\lambda}|^2}{\sum_{\eta}\sum_{\bar{\eta}}\sum_{\lambda=\pm 1}|s_{\eta,\lambda}^{\bar{\eta},-\lambda}|^2+|s_{\eta,\lambda}^{\bar{\eta},\lambda}|^2}.
\end{equation}
Note that the symbolic sums in $\eta$ and $\bar{\eta}$ may contain integrals for continuous labels like linear momentum and/or sums for discrete labels like angular momentum. 

The measure defined in Eq. (\ref{eq:cancelD}) ranges from 0 to 1. Zero corresponds to complete helicity preservation (i.e. duality symmetry of the scatterer) and 1 to a scatterer that completely flips the helicity of the incident field. Importantly, $\cancel{D}$ is basis independent.

In the context of the macroscopic Maxwell's equations, a scatterer made of a material characterized by relative electric permittivity and magnetic permeability $(\epsilon_s,\mu_s)$ embedded in a background with material properties $(\epsilon,\mu)$ has duality symmetry, and therefore preserves helicity ($\cancel{D}=0$), if and only if \cite{FerCor2012p}:

\begin{equation}
\label{eq:duality_macro}
\boxed{\frac{\epsilon_s}{\mu_s}=\frac{\epsilon}{\mu}}.
\end{equation}

For optical activity, it would be desirable to build a chiral structure with materials meeting Eq. (\ref{eq:duality_macro}). The problem is that these kind of materials can be obtained for radio frequencies \cite{Schubring2004,sengupta2000}, but not for other frequency ranges like the optical one. While dual symmetric materials are not available for most frequencies, the situation is different for small scatterers in the dipolar approximation. A small scatterer is considered ``dipolar'' if its response to an electromagnetic field can be described to good approximation by just the electric $\ed$ and magnetic $\md$ moments induced by the incident field $(\EE(\rrp),\HH(\rrp))$ at the position $\rrp$ of the small scatterer. For a dipolar scatterer to be dual symmetric, i.e to preserve the helicity of the incident field, its polarizability tensor $P$ 
\begin{equation}
\label{eq:M}
\begin{bmatrix}\ed\\\md\end{bmatrix}=P\begin{bmatrix}\EE(\rrp)\\\HH(\rrp)\end{bmatrix}=
\begin{bmatrix}\MdE & \MdH\\\MmE & \MmH\end{bmatrix}
\begin{bmatrix}\EE(\rrp)\\\HH(\rrp)\end{bmatrix},
\end{equation}
must meet \cite{FerCor2013}
\begin{equation}
\label{eq:dual_dipolar}
\boxed{\MdE=\epsilon\MmH,\ \MmE=-\frac{\MdH}{\mu}}.
\end{equation}
When a field of well defined helicity, i.e. which has zero content of one of the two helicities, interacts with a dipolar object whose response meets Eq. (\ref{eq:dual_dipolar}), the induced electric and magnetic dipoles have a fixed relationship 
\begin{equation}
	\label{eq:dm}
\ed=\pm\frac{ i}{c}\md,
\end{equation}
where the $\pm$ corresponds to the two possible helicities of the incident field. It can be shown that the combined field radiated by the induced dipoles is of well defined helicity, and equal to the one of the incident field \cite{FerCor2013}. Therefore, when the incident field contains the two helicities, the interaction with such a scatterer does not couple them.

There are realistic scatterers that meet Eq. (\ref{eq:dual_dipolar}) even at microwave \cite{Geffrin2012} and optical frequencies \cite{Person2013}. In these works, properly designed dielectric spheres and cylinders have been empirically shown to exhibit zero backscattering, which is achieved by dual symmetric objects with discrete rotational symmetries of degree higher than two \cite{FerCor2013c}. The duality properties of dielectric spheres have been studied by Zambrana {\em et al.} in Ref. \cite{Zambrana2013b}. Their work shows that, by adequately choosing its radius $r$ and relative electric permittivity $\epsilon_s$, a dielectric sphere can be made dual symmetric in the dipolar approximation (\ref{eq:dual_dipolar}). For a sphere, the polarizability tensor is completely determined by the dimensionless Mie scattering coefficients (see e.g. \chooseformat{\onlinecite{Chap. 3.4}{Wheeler2010PHDTH}}{\cite[Chap. 3.4]{Wheeler2010PHDTH}}), and dipolar duality is equivalent to the equality of the first order electric and magnetic Mie scattering coefficients: $a_1=b_1$. Particles with such properties have attracted attention, particularly in the context of zero backscattering and the Kerker conditions \cite{Geffrin2012,Zambrana2013}.

\begin{figure*}[ht]

	\begin{tabular}{cc}

		\subfloat[]{\label{fig:dbreak}
		\chooseformat{\includegraphics[height=7.5cm]{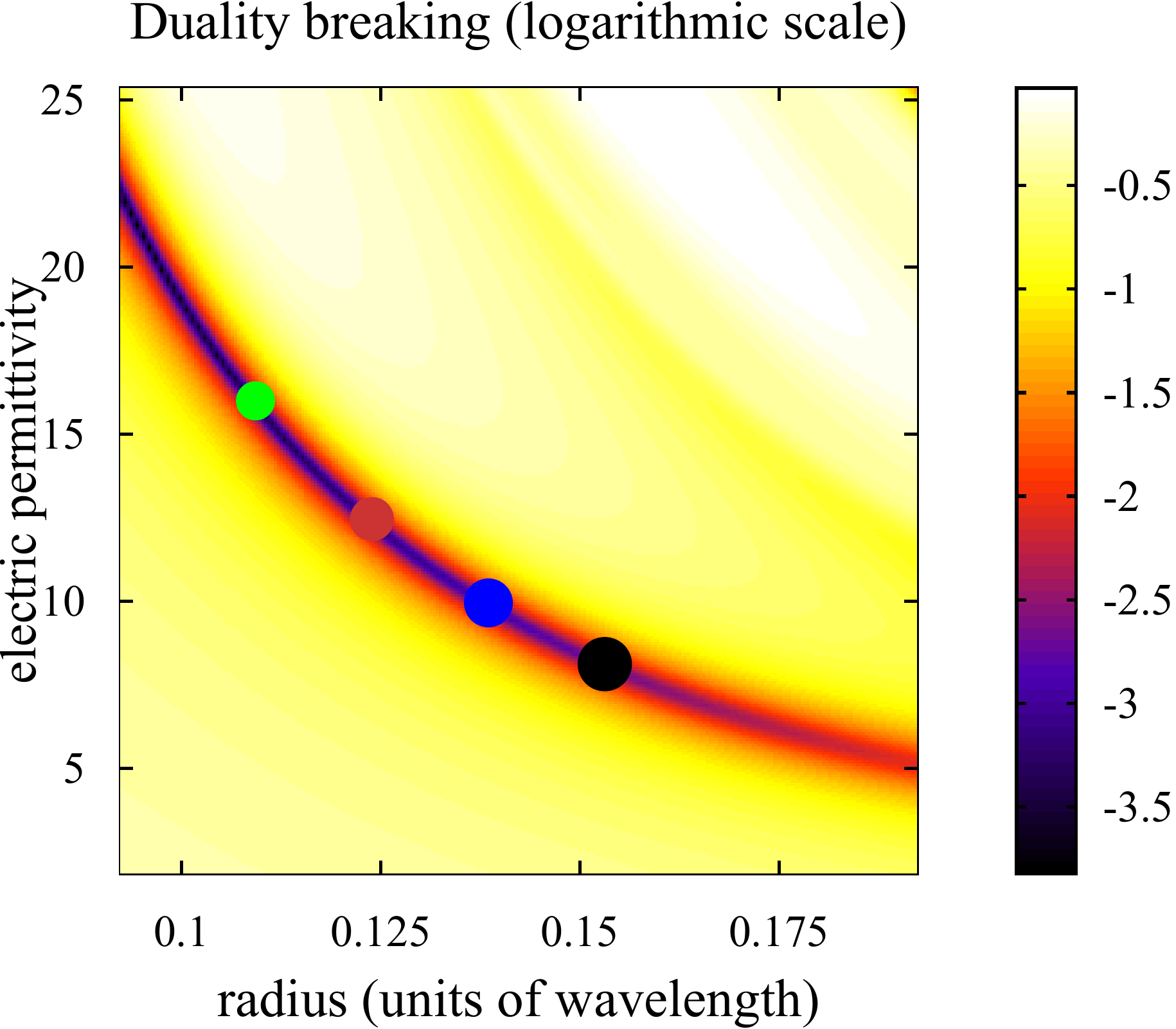}}
		{\includegraphics[height=7.5cm]{dbreak.pdf}}
		}
		\begin{picture}(0,0)
			\setlength{\fboxrule}{0.75pt}
			\setlength{\fboxsep}{0pt}
			\put(-125,25){
				\chooseformat{\raisebox{4.15cm}{\framebox{\includegraphics[width=2cm]{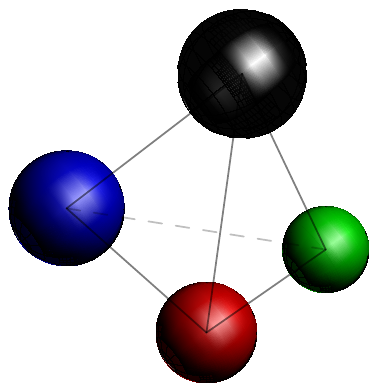}}}}
				{\raisebox{3.75cm}{\framebox[1.01\width][c]{\includegraphics[width=1.75cm]{structure.pdf}}}}		
				}
\end{picture}
			&
	\hspace{0cm}\subfloat[]{\label{fig:a1b1}
	
	\chooseformat{\includegraphics[height=7cm]{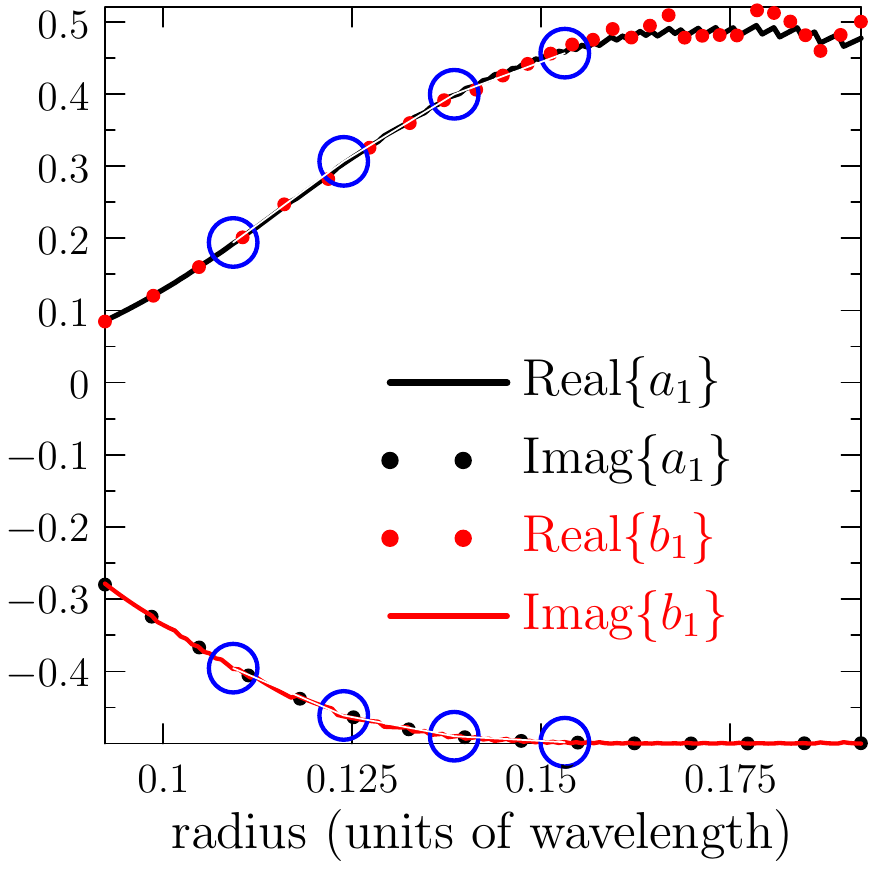}}{\includegraphics[height=7cm]{a1b1}}}\\

	\end{tabular}

	\caption{\label{fig:dualbreak} Inset panel (a) : A tetrahedral arrangement of four different spheres is a chiral object. If the spheres are chosen to be dipolarly dual symmetric, we expect that the structure preserves helicity to good approximation for many incident/scattered directions. Such an object is designed using the known necessary conditions for optical activity in general scattering directions. Panel (a): Relative helicity change by an individual dielectric sphere immersed in a medium with permittivity equal to $(1.3)^2$. Equation (\ref{eq:cancelD}) defines the relative helicity change of an arbitrary scatterer as the ratio between the sum of the modulus square of all the helicity flipping scattering coefficients and the sum of the modulus square of all the scattering coefficients, which ranges from 0 to 1. The plot shows a region of very small helicity change. We design a chiral and approximately dual tetrahedron made with four spheres with parameters indicated by the white circles, and compare it to two other structures: A chiral and non-dual tetrahedron made with four spheres with parameters indicated by the black circles, and a chiral and perfectly dual tetrahedron made with materials meeting $\epsilon=\mu$. Panel (b): Electric $a_1$ and magnetic $b_1$ dipolar Mie scattering coefficients as a function of the sphere radius for the choice of relative electric permittivity that minimizes the helicity change for each radius. As expected, such choice is very close to the dipolar duality condition for spheres ($a_1=b_1$). The blue circles mark the dipolar coefficient values of the four spheres in the approximately dual tetrahedron.} 
\end{figure*}

The dipolar approximation ignores higher multipolar terms which will, in general, break helicity preservation resulting in $\cancel{D}\neq0$. In the case of a sphere, such duality breaking terms can be analytically computed using Mie theory \chooseformat{\onlinecite{Sec. 2}{Zambrana2013b}}{\cite[Sec. 2]{Zambrana2013b}}.  Figure \ref{fig:dbreak} shows $\cancel{D}$ for a sphere as a function of radius and electric permittivity. The figure shows a region of the parameter space $(r,\epsilon_s)$ where the total relative helicity change due to all multipolar terms is quite small ($\cancel{D}\approx 10^{-3}$). Pairs of geometrical and material properties for approximately dual spheres can be extracted from this region. The calculations shown in Fig. \ref{fig:dbreak} assume that the medium surrounding the spheres has $\epsilon=(1.3)^2$. For a surrounding medium with a different relative electric permittivity $\bar{\epsilon}$, the same results will be obtained by changing $\epsilon_s$ to $\epsilon_s\bar{\epsilon}/(1.3)^2$ while simultaneously changing the radius $r$ to $1.3r/\sqrt{\bar{\epsilon}}$.
  
We may use several dipolarly dual spheres to build composite scatterers which preserve helicity to good approximation. For optical activity, we also need the breaking of mirror symmetries, which can be achieved by assembling spheres into a chiral configuration, considered then the structure of interest. One possible configuration is a tetrahedral arrangement of four dipolarly dual spheres, like the one in the inset of Fig. \ref{fig:dbreak}. If the four spheres in the tetrahedron have different electromagnetic responses, the arrangement in the inset of Fig. \ref{fig:dbreak} is chiral. Figure \ref{fig:a1b1} shows some of the available range of $a_1(b_1)$ coefficients for dipolarly dual spheres, which largely determine their response. We have selected four spheres with parameters indicated by the white circles in Fig. \ref{fig:dbreak} and blue circles in Fig. \ref{fig:a1b1}.

Due to its chirality, the tetrahedral arrangement meets the condition of breaking all mirror symmetries. As previously discussed, the duality condition is only approximately met. In order to gauge the effect of this approximation we will compare the design with two other different tetrahedral structures. One is composed of four exactly dual (magnetic) spheres of different sizes made with materials meeting Eq. (\ref{eq:duality_macro}): $\epsilon_1=\mu_1=16.00,\ \epsilon_2=\mu_2=12.46,\ \epsilon_3=\mu_3=9.95$, and $\epsilon_4=\mu_4=9.12$. The other one is composed of dielectric spheres whose parameters set them far away form the dipolar condition of Eq. (\ref{eq:dual_dipolar}). They are marked by black circles in  Fig. \ref{fig:dbreak}. 

Table \ref{tab:structure} contains the specification of the three structures, and includes the helicity change for each isolated sphere. For the sake of conciseness, we will refer to the dual structure as DS, to the approximately dual structure as ADS and to the structure made with spheres that severely break duality symmetry as NDS, which stands for non-dual structure. The permittivities of the spheres in the proposed ADS are available at optical frequencies. The intrinsically magnetic materials in the DS are not. The DS is used in this article as an ideal reference for comparison purposes.

\begin{table}[h!]
	{\footnotesize
\begin{center}\begin{tabular}{l|c|ccc|ccccc} \toprule
{\bf Positions} & {\bf Radii} & &$\boldsymbol\epsilon\boldsymbol/\boldsymbol\mu$&& \multicolumn{3}{c}{{\bf Helicity change}}  \\

				& & DS & ADS & NDS&DS&  ADS & NDS\\\midrule
		$(1,1,1)\frac{0.32}{\sqrt{3}}$&0.110& 16.00/16.00 & 16.00/1 &11.20/1&0&3e-4&0.186\\
 $(1,\shortminus1,\shortminus1)\frac{0.32}{\sqrt{3}}$&0.124& 12.46/12.46&12.46/1&16.19/1&0&6e-4&0.421\\
 $(\shortminus1,1,\shortminus1)\frac{0.32}{\sqrt{3}}$&0.138& 9.95/9.95&9.95/1&6.97/1&0&1.0e-3&0.157\\
		$(\shortminus1,\shortminus1,1)\frac{0.32}{\sqrt{3}}$&0.153& 9.12/9.12&9.12/1&10.56/1&0&2.0e-3&0.209\\\bottomrule
\end{tabular}\end{center}
}
\caption{\label{tab:structure}Specification of the spheres in each of the three structures analyzed in the article: DS stands for dual structure, ADS for approximately dual structure and NDS for non-dual structure. The different columns show, respectively from left to right, the positions of the spheres, their radii, their relative electric permittivity and magnetic permeability, and the total helicity change for each individual sphere (see Fig. \ref{fig:dbreak}). The positions and radii are in units of wavelengths.} 
\end{table}

The scattering matrix of each tetrahedron can be numerically computed given the positions, radius, and material properties of its composing spheres \cite{Mackowski1996,Xu1995,Muhlig2010,Fruhnert2014}. We compute the scattering matrices of the tetrahedrons to multipolar order 9, which achieves sufficient convergence. These matrices encode all the information about the electromagnetic properties of each structure. Using them, we can calculate the total scattering cross sections, which are 3.99 for the ADS, 4.46 for the NDS and 4.08 for the DS, in units of the individual scattering cross section of the sphere with parameters ($r=0.153$, $\epsilon=9.12$, $\mu=1$). We can also compute their relative helicity change $\cancel{D}$, which ranges from zero to one. The results are $0.0012$ for the ADS, $0.2731$ for the NDS and, of course, zero for the dual structure. This confirms that the ADS preserves helicity to good approximation. 

We now analyze the optical activity of each structure. The next section describes the analysis methodology.

We note that the use of the wavelength as the unit of length for the radii and positions of the spheres in the tetrahedrons renders the analysis and results independent of the specific wavelength. In particular, the values of $\cancel{D}$ for a sphere (Fig. \ref{fig:dbreak}), the values of the total scattering cross sections and $\cancel{D}$ for the tetrahedrons, and the optical activity results contained in the next sections are wavelength independent. A change of wavelength, i.e, a change of the unit of length, re-scales all spatial dimensions accordingly. The results for the re-scaled structure will be the same as before (assuming the same material parameters). The ability to realize the structure at a given working wavelength depends on whether there exist materials with the adequate electric permittivity at that wavelength. 
 
\section{Analysis methodology}\label{sec:an}
Given a scattering matrix in the basis of multipolar fields of well defined parity, the computation of the subscattering matrix between two incident/scattered plane waves $\p/\pbar$ in the helicity basis is straightforward (\chooseformat{\onlinecite{Eqs. 11.4-6, 8.4-(9,10)}{Tung1985}}{\cite[Eqs. 11.4-6, 8.4-(9,10)]{Tung1985}}). After such computation we obtain a 2 by 2 subscattering matrix for each specified pair $\p/\pbar$ 
\begin{equation}
S_\p^\pbar=\begin{bmatrix}\aaa & \bbb\\\ccc&\ddd\\\end{bmatrix},
\end{equation}
where $\aaa$ and $\ddd$ are helicity preserving coefficients for positive and negative helicity, respectively, and $\bbb$ and $\ccc$ are helicity changing coefficients. These subscattering matrices encode the polarization change in the helicity basis and correspond to the Jones matrices in such basis. Their consideration simplifies the analysis of optical activity.

For an incident linear polarization with angle \footnote{The way to measure polarization angles is as follows: For a plane wave with its momentum aligned with the z-axis, the zero of linear polarization angle is assigned to the x-axis. For an arbitrary plane wave with momentum $\p$, the corresponding zero reference is obtained rotating the x-axis by the same rotation that brings $\zhat$ into $\p/|\p|$.} $\alpha_\p$, the output polarization components are 

\begin{equation}
\label{eq:tx}
	\begin{split}
\begin{bmatrix}{\Fp}\\{\Fm}\end{bmatrix}
	&=
\begin{bmatrix}
\aaa & \bbb\\\ccc&\ddd\\
\end{bmatrix}
\begin{bmatrix}\exp(-i\alpha_\p)/\sqrt{2} \\\exp(i\alpha_\p)/\sqrt{2}\end{bmatrix}\\
	&=\frac{1}{\sqrt{2}}
\begin{bmatrix}\aaa\exp (-i\alpha_\p)+\bbb\exp (i\alpha_\p)\\\ccc\exp (-i\alpha_\p)+\ddd\exp (i\alpha_\p) \end{bmatrix}.
\end{split}
\end{equation}

The angle of the major axis of the output polarization ellipse is 
\begin{equation}
	\label{eq:theta}
	\theta_\p^{\pbar}=\frac{1}{2}\arg{\left({\Fm}{\Fp}^*\right)}. 
\end{equation}
We define the rotation angle $\beta_\p^\pbar$ as the difference between $\theta_\p^{\pbar}$ and $\alpha_\p$:
\begin{equation}
	\label{eq:beta}
	\beta_\p^{\pbar}=\theta_\p^{\pbar}-\alpha_\p.
\end{equation}

The ellipticity of the output polarization can be defined as  
\begin{equation}
	\label{eq:eta}
	\eta_\p^\pbar=\frac{\left(A_\p^\pbar\right)^2-\left(B_\p^\pbar\right)^2}{\left(A_\p^\pbar\right)^2+\left(B_\p^\pbar\right)^2},
\end{equation}
where $A_\p^{\pbar}$ and $B_\p^{\pbar}$ are the major and minor axis of the ellipse. This measure of ellipticity takes the extreme values of 1 for a linearly polarized output and zero for a circularly polarized output.

It can be shown that the ellipticity $\eta_\p^\pbar$ can be computed as
\begin{equation}
	\label{eq:eta2}
	\eta_\p^\pbar=\frac{2|{\Fm}{\Fp}^*|}{|\Fp|^2+|\Fm|^2}.
\end{equation}

In general, $\beta_\p^{\pbar}$ and $\eta_\p^\pbar$ depend on $\alpha_\p$. By varying $\alpha_\p$ in Eqs. (\ref{eq:beta}) and (\ref{eq:eta}), we obtain the functions $\beta_\p^\pbar(\alpha_\p)$ and $\eta_\p^\pbar(\alpha_\p)$. Crucially, it can be shown that $\beta_\p^\pbar(\alpha_\p)$ is independent of $\alpha_\p$ if and only if $\bbb=\ccc=0$, i.e. when helicity is preserved \cite{FerCor2012c}. In such case, $\eta_\p^\pbar(\alpha_\p)$ is independent of $\alpha_\p$ as well. \chooseformat{The Supp. Info.}{Appendix \ref{app:optact}} contains these brief derivations, complementing those in \cite{FerCor2012c} with the treatment of ellipticity.

\begin{figure}[h!]
	\chooseformat{
	\includegraphics[width=8cm]{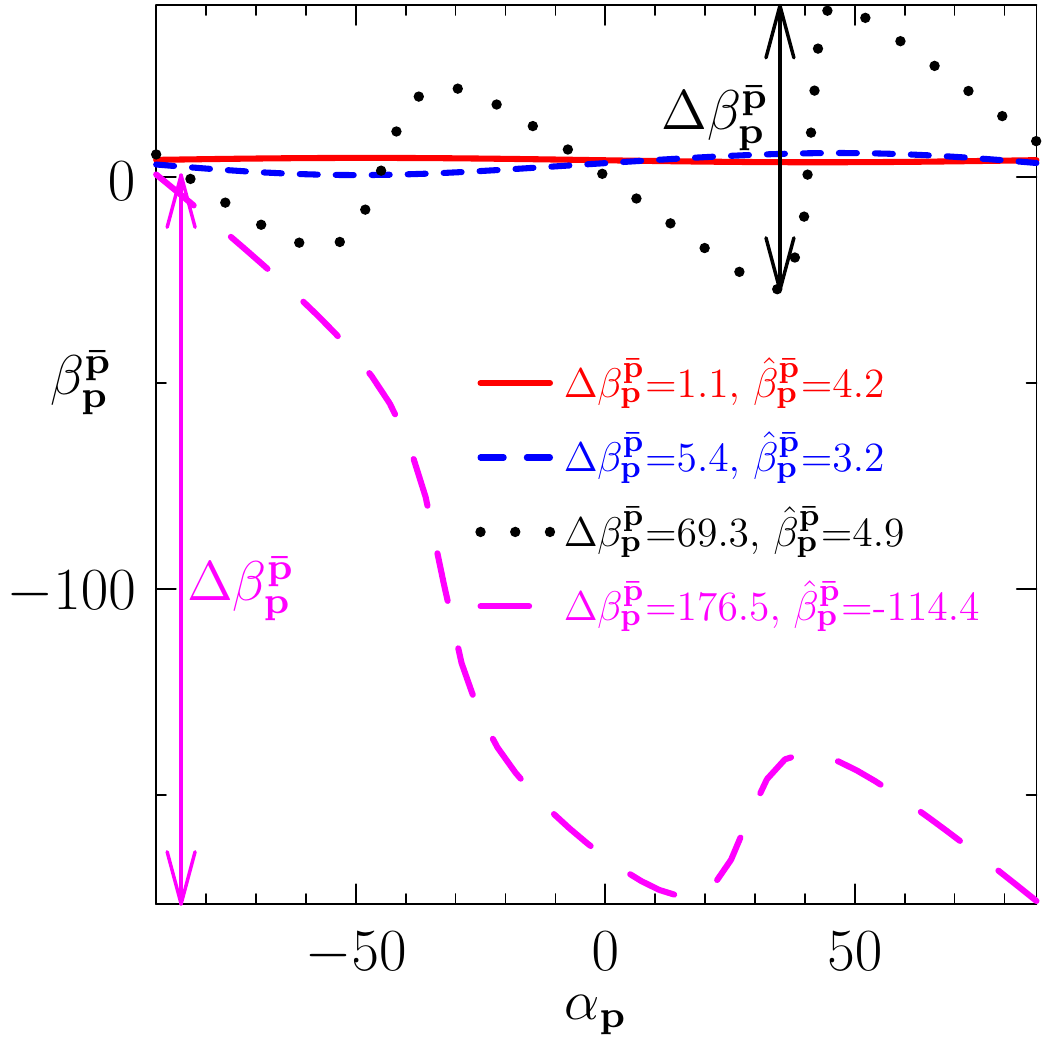}}
	{\includegraphics[width=8cm]{fig_beta_deltabeta.pdf}}

	\caption{\label{fig:betadelta} Output rotation angle $\beta_\p^\pbar$ as a function of the incident polarization angle $\alpha_\p^\pbar$ for four cases occurring in the analyzed structures. All angular quantities are in degrees. Optical activity implies a flat line ($\partial\beta_\p^{\pbar}/\partial\alpha_\p^{\pbar}=0$). For the continuous red and short-dashed blue cases, where the peak to peak variation of $\beta_\p^\pbar$ is small, the polarization transformation may be considered a rotation by a fixed angle, akin to optical activity (see Fig. \ref{fig:poltrans}). When $\Deltabeta$ is large, as in the dotted black and long-dashed magenta examples in the figure, such consideration is much less adequate. The $\betahat$ in the legend correspond to the average rotation in each case. The first two cases have been extracted from the ADS data. The polar and azimuthal angles defining their $\p/\pbar$ vectors are $(13.5,0.0)/(40.5,180.0)$ and $(13.5,279.0)/(148.5,216.0)$. The second two cases have been extracted from the NDS data. They correspond to the pairs $(126.0,9.00)/(67.5,117.0)$ and $(9.0,351.0)/(108.0,243.0)$. The four exemplary cases have been chosen to clearly illustrate small and large $\Deltabeta$. The sphere positions in Table 1 fix the orientation of the structures with respect to the coordinate axes where angles are measured.}
\end{figure}

Consequently, optical activity for the $\p/\pbar$ directions requires $\bbb=\ccc=0$. In the DS, duality symmetry implies $\bbb=\ccc=0$ for all $\p/\pbar$, and $\beta_\p^\pbar$ and $\eta_\p^\pbar$ are independent of $\alpha_\p$ for all $\p/\pbar$. We may expect the ADS to deviate slightly from this optimal situation and the NDS to have a large deviation. To measure these deviations we define $\Deltabeta$ and $\Deltaeta$ as the length of the range covered by $\beta_\p^\pbar(\alpha_\p)$ and $\eta_\p^\pbar(\alpha_\p)$ as $\alpha_\p$ varies, i.e. their peak to peak variation. We also define the average rotation $\betahat$ and ellipticity $\etahat$ as the average of $\beta_\p^\pbar(\alpha_\p)$ and $\eta_\p^\pbar(\alpha_\p)$ as $\alpha_\p$ varies. Figure \ref{fig:betadelta} contains examples of different $\beta_\p^\pbar(\alpha_\p)$ behaviors taken from actual data. The consideration of the polarization transformation as optical activity becomes less adequate as $\Deltabeta$ increases.

\section{Discussion of the numerical results}\label{sec:dis}
Following the methodology of the previous section, we compute $\betahat,\etahat,\Deltabeta$ and $\Deltaeta$ for all possible pairs $\p/\pbar$ on a double grid sampling the continuum of spatial directions in steps of 4.5 degrees in the polar coordinate and 9 degrees in the azimuthal coordinate. The total number of pairs is $(180/4.5\times360/9)^2=(1600)^2$. All angular quantities are in degrees.

\begin{figure}[ht!]

\includegraphics[width=8cm]{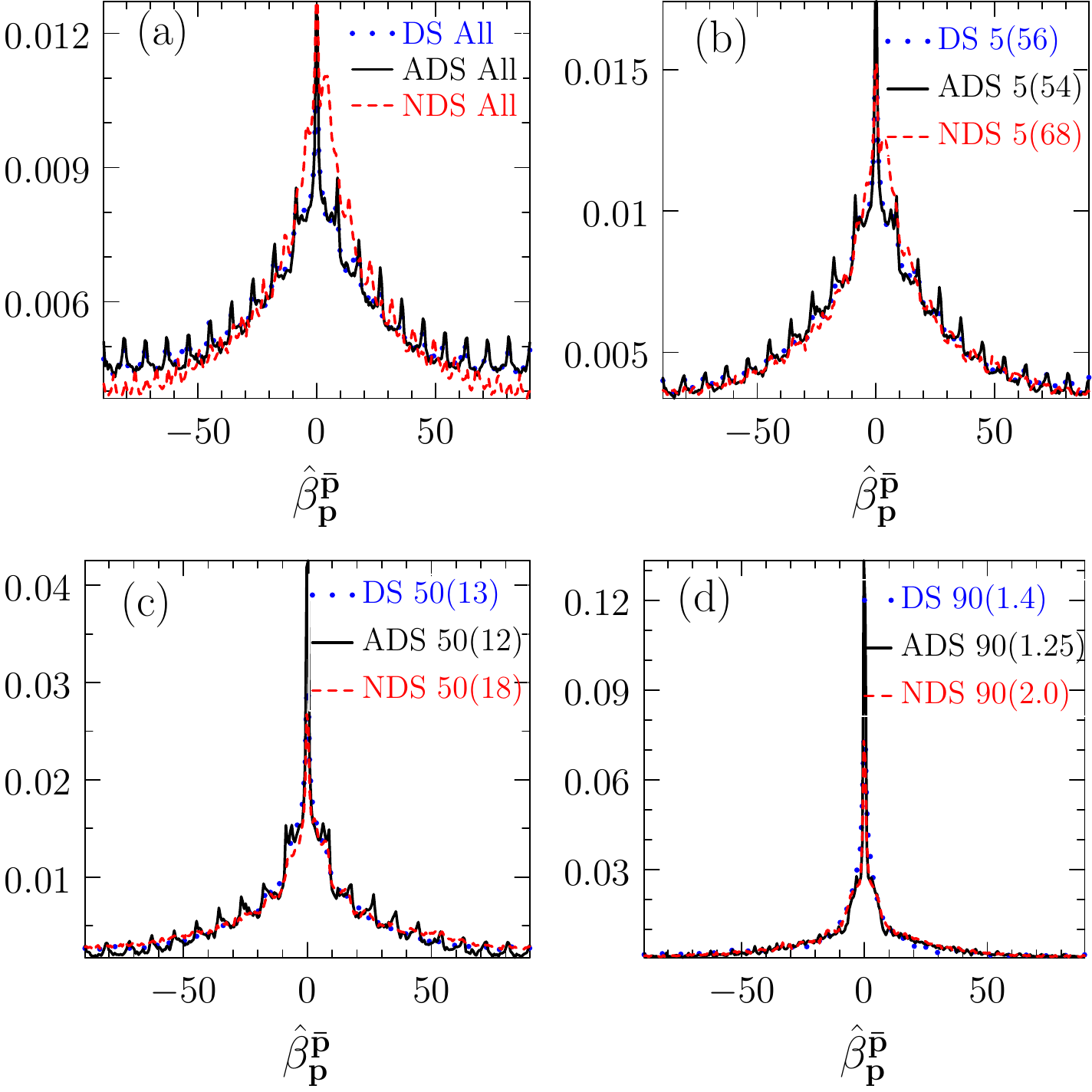}
\caption{\label{fig:kpi} Panels (a)-(d): Histograms of $\betahat$, the rotation angle averaged over the incident linear polarization angle $\alpha_\p^\pbar$ for fixed $\p/\pbar$ for the DS, ADS and NDS. All angular quantities are in degrees. The figures show aggregated data for all the $\p/\pbar$ pairs in a double grid of incident/scattered directions sampled in steps of 4.5 degrees in the polar coordinate and 9 degrees in the azimuthal coordinate. The total number of pairs is $(180/4.5\times360/9)^2=(1600)^2$. Figure (a) contains all data. In Figs. (b), (c), and (d) the weaker scattering pairs $\p/\pbar$ that together add up to 5, 50 and 90 \% of the total scattering cross section have been excluded. The X(Y) notation means that for a scattering threshold at X\%, Y\% of the total $\p/\pbar$ pairs are kept. The results are discussed in the text.}
\end{figure}

In Fig. \ref{fig:kpi}, each plot shows the statistics from a reduced set of $\p/\pbar$ pairs containing the stronger scattering pairs. Figure \ref{fig:kpi}a contains all the cases. In Fig. \ref{fig:kpi}b the weaker $\p/\pbar$ pairs are excluded, which together add up to 5\% of the total scattering cross section. In Figs. \ref{fig:kpi}c,b this threshold is 50\% and 90\%, respectively. In other words, Fig. \ref{fig:kpi}d shows the data for the stronger scatterer pairs which, together, add up to 10\% of the total scattering cross-section; and similarly for the other plots. The percentages of $\p/\pbar$ cases kept by the 5-50-90 filtering is 54-12-1.25 for the ADS, 68-18-2.0 for the NDS and 56-13-1.4 for the DS. The results of Figs. \ref{fig:kpi}a-d show that the three structures have non-zero average rotation angles $\betahat$ for many scattering pairs. Even though the histograms change as the threshold increases, for example, some features disappear, the dispersion of $\betahat$ values is still quite large even in Fig. \ref{fig:kpi}d. The data shows that the non-zero $\betahat$ values are not limited to weakly scattering $\p/\pbar$ cases. The occurrence of different $\betahat$ values is consistent with the fact that there is no reason to expect a single $\betahat$ value for all $\p/\pbar$ in any of the three structures.

Figures S3a-d in \chooseformat{the Supp. Info.}{appendix \ref{app:eta}} contain the histograms for the average ellipticity $\etahat$. The values of $\etahat$ are shifted towards large values and this shift is more pronounced as the scattering threshold increases. This indicates that, upon incident linear polarization, the outputs are much closer to being linearly polarized ($\eta=1$) than circularly polarized ($\eta=0$). 

\begin{figure*}[ht!]
\includegraphics[width=15cm]{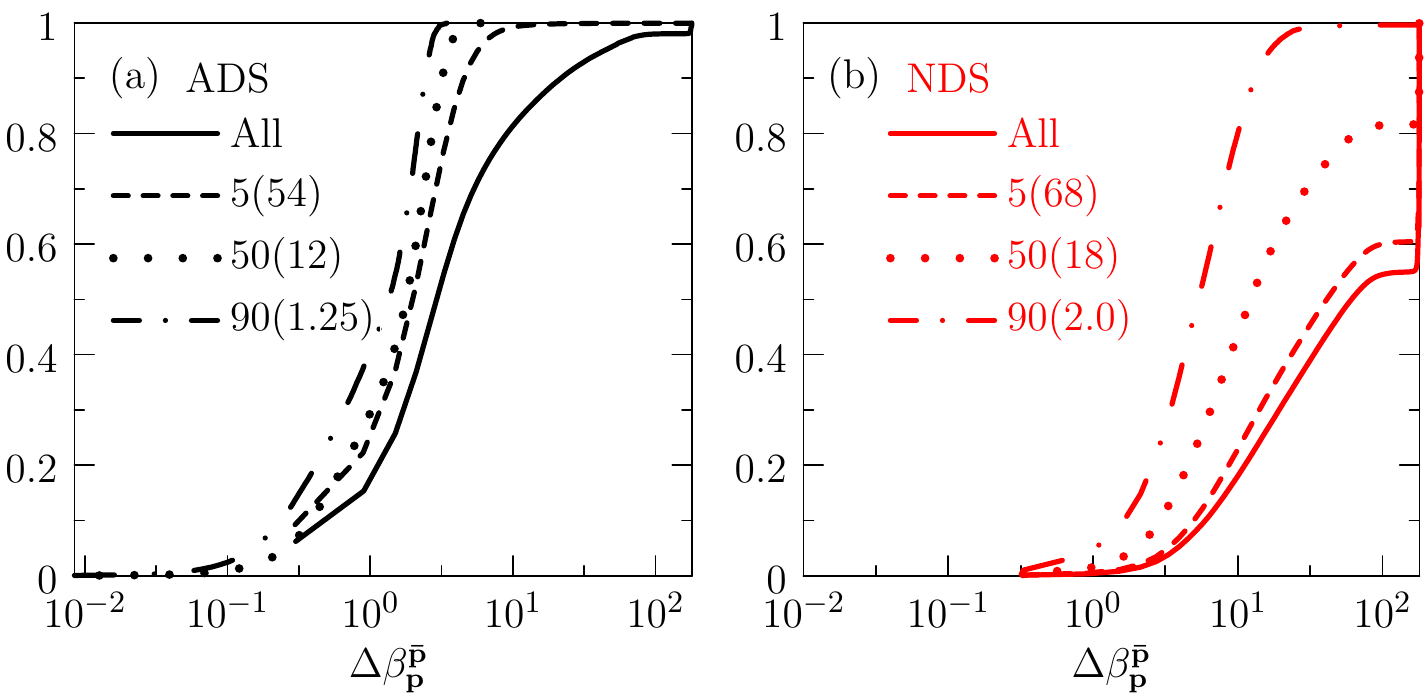}
\caption{\label{fig:deltabetafig} Panels (a,b): Cumulative histograms of $\Deltabeta$, the peak to peak variations of the rotation angle for the ADS and NDS, respectively. The plots contain data for the different settings of the scattering threshold (see the text or the legend of Fig. \ref{fig:kpi}). Note the logarithmic scale in the horizontal axis. The values of $\Deltabeta$ are significantly smaller in the ADS (a) than in the NDS (b). The results are further discussed in the text.}
\end{figure*}

The statistics of $\Deltabeta$ in Figs. \ref{fig:deltabetafig}a,b allow us to judge whether the non-zero average rotation angles from Figs. \ref{fig:kpi}a-d {\em can be meaningfully considered as optical activity, or not} according to the previous discussions. The ideal result is a step function rising at $\Deltabeta=0$. This is achieved by the DS (not shown in the figures). For the ADS and NDS structures, the polarization transformations that occur in the scattering direction pairs $\p/\pbar$ are closer to optical activity for smaller values of $\Deltabeta$ and sharper rises of the cumulative histograms. The results clearly show the difference between the ADS (Fig. \ref{fig:deltabetafig}a) and the NDS (Fig. \ref{fig:deltabetafig}b) in this respect. For the ADS, 95\% of the pairs have a $\Deltabeta$ smaller than 43.5, 10.3, 3.5 and 2.6 degrees, respectively as the threshold increases (note the logarithmic scale in the horizontal axis). For the NDS, the jumps in the rightmost bins of the plots indicate that a portion of the $\p/\pbar$ pairs have $\Deltabeta>180$ degrees, which we set to $\Deltabeta=180$ degrees in the statistics. These cases are similar to the magenta case in Fig. \ref{fig:betadelta}. For the first three thresholds (0-5-50), the 95\%-ile of $\Deltabeta$ lies beyond $\Deltabeta=180$ in the NDS. When the threshold discards the weaker scatterers adding up to 90\% of the total scattering, the 95\%-ile of $\Deltabeta$ is 16.7 degrees, compared 2.6 for the ADS. Figs. \ref{fig:deltabetafig}a,b show that the NDS exhibits a huge deviation with respect to the ideal case, and that the ADS approaches the ideal optical activity performance of the DS reasonably well. Figures S3e,f in \chooseformat{the Supp. Info.}{appendix \ref{app:eta}} show the cumulative histograms for the variation in ellipticity $\Deltaeta$, which are in line with this conclusion. The 95\%-iles for $\Deltaeta$ in the ADS structure are 0.46, 0.14, 0.03 and 0.01; and in the NDS 0.91, 0.91, 0.89 and 0.33. The ideal result is $\Deltaeta=0$, achieved by the DS.

The results match the expectations. In sharp contrast to the NDS case, the polarization transformation in many of the scattering pairs of the ADS meets the definition of optical activity to a very good approximation. For example, let us take the data for all the $\p/\pbar$ pairs adding up to the 95\% of the total scattering (which is 54\% of the total number of pairs); only 5\% of those cases have a $\Deltabeta$ larger than 5.5 degrees. This is quite different in the NDS case. For the same scattering threshold (which keeps 68\% of the total number of pairs), $\Deltabeta$ is larger than 5.5 degrees for 89\%. 

\begin{figure}[ht!]

\includegraphics[width=8cm]{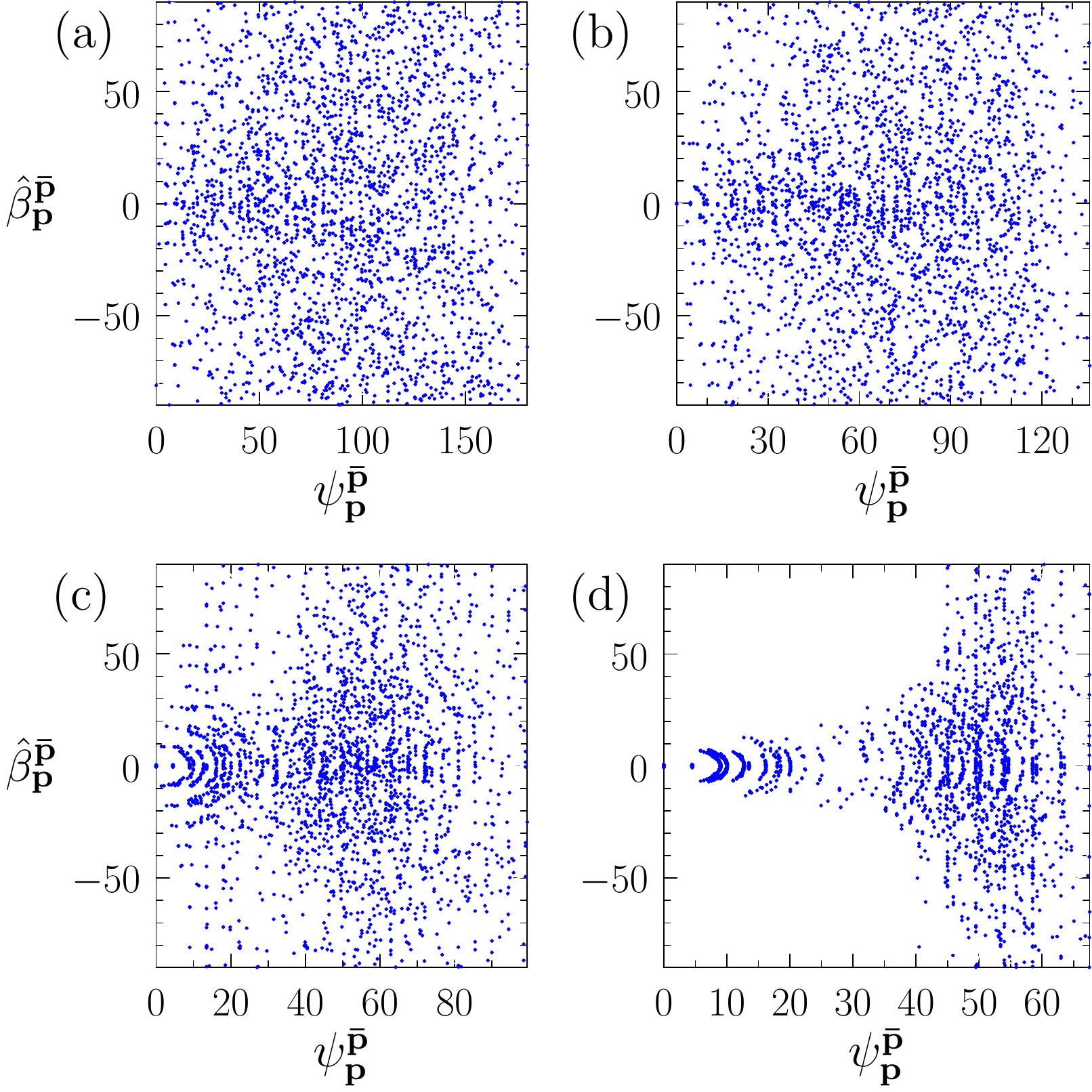}

\caption{\label{fig:multidirectional} Scatter plots of average rotation $\betahat$ versus the angle formed by the incident and scattered directions $\psippbar=\textrm{angle}(\p,\pbar)$ for the ADS. All angular quantities are in degrees. The (a)-(b)-(c)-(d) panels contain the data for the four settings of the scattering threshold (see the text or the legend of Fig. \ref{fig:kpi}). For the sake of clarity, the number of points in each plot is limited to 2500 by random down sampling. The plots show that the ADS produces optical rotation in general scattering directions.}
\end{figure}

Figure \ref{fig:deltabetafig}a shows that the polarization transformations effected by the ADS can, in many cases, be meaningfully considered optical activity. Finally, Fig. \ref{fig:multidirectional} shows that the ADS produces optical rotation in forward and non-forward scattering directions. The four figures contain the scatter plots of the average rotations $\betahat$ versus the angle formed by the incident and scattered directions $\psippbar=\textrm{angle}(\p,\pbar)$ for the four settings of the scattering threshold. The observed range of optical rotation angles is maximal ($\pm 90$ deg.) for many scattering angles $\psippbar$. In forward scattering, the range is reduced to $\pm 0.5$ degrees when the scattering threshold is 5, 50 or 90. We attribute this effect to the already discussed fact that, while any other scattering direction requires the breaking of a single plane of symmetry, optical activity in forward scattering requires the breaking of all mirror planes of symmetry containing the optical axis, resulting in more chances that the symmetry is only weakly broken for one of them.

These results confirm and highlight the importance of helicity preservation in optical activity and show that this design requirement can be addressed with the use of approximately dual structures, achieving optical activity in general scattering directions. 

\section{Concluding remarks}\label{sec:con}
In summary, optical activity in general scattering directions can be achieved by chiral structures that, additionally, have electromagnetic duality symmetry. At frequency ranges where dual symmetric materials are not available, like the optical one, chiral and approximately dual structures can be devised using small scatterers whose dipolar response is dual symmetric. We have shown that this design strategy allows to address the requirement of helicity preservation, necessary for optical activity, and results in structures that exhibit optical activity in general scattering directions. 

Structures such as the one studied in this article are suitable building blocks for macroscopic objects exhibiting optical activity in general scattering directions. For example, a two dimensional array of copies of the presented structure should exhibit optical activity in both reflection and transmission at oblique incidence.

Electromagnetically small single objects that are chiral and dipolarly dual exist at microwave frequencies \cite{Semchenko2009}. Obtaining them at optical frequencies would offer an alternative to achieving chirality by the spatial arrangement of non chiral objects.

We believe that the consideration of the electromagnetic duality symmetry is a valuable addition to the research in optical activity, for both its fundamental and practical sides.

\begin{acknowledgments}
	This work was supported by the German Science Foundation within project RO 3640/3-1.
\end{acknowledgments}
\clearpage
\bibliographystyle{ifcbst}

\clearpage
\appendix
\section{Helicity and duality}\label{app:heldual}
The helicity operator is the projection of the total angular momentum vector operator onto the linear momentum direction
\begin{equation}
\Lambda=\frac{\mathbf{J}\cdot{\mathbf{P}}}{|\mathbf{P}|}.
\end{equation}

For classical electromagnetic fields in the complex notation\footnote{The choice of complex fields with only positive frequencies $\omega>0$ has the important physical implication of choosing only positive energies.}, helicity has two possible eigenvalues $\pm1$. The eigenstates of helicity are the linear combinations $\mathbf{E}\pm i Z\mathbf{H}$, with $Z$ a reference impedance, so that
\begin{equation}
\Lambda \left(\mathbf{E}\pm i Z\mathbf{H}\right)=\pm \left(\mathbf{E}\pm i Z\mathbf{H}\right).
\end{equation}

\begin{figure}[h!]
	\includegraphics[width=\linewidth]{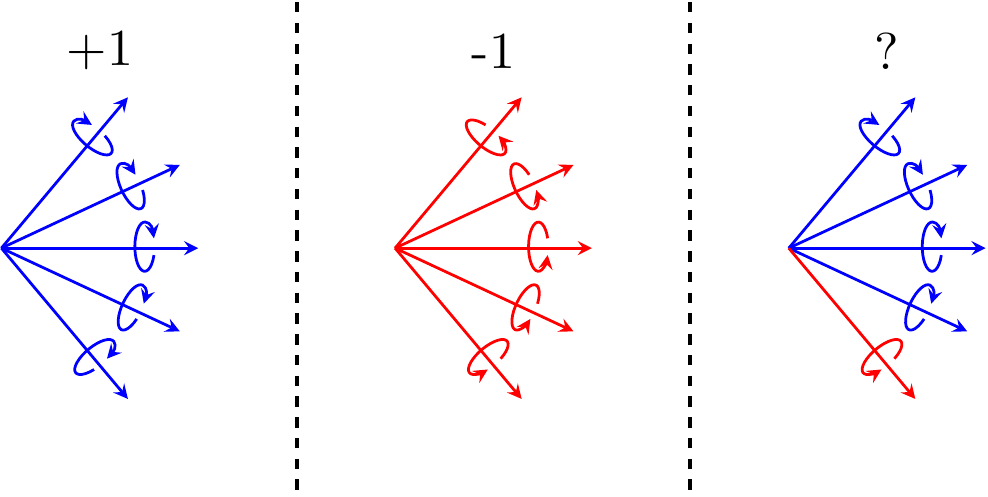}
\caption{\label{fig:helicity}A field composed by the superposition of five plane waves has definite helicity equal to one if, with respect to their momentum vectors, all the plane waves are left hand polarized (left part), equal to minus one if they are all right hand polarized (central part), and does not have a definite helicity if all the plane waves do not have the same polarization handedness (right part).}
\end{figure}

One intuitive way of understanding helicity is as the polarization handedness in momentum space. Figure \ref{fig:helicity} illustrates this.

As an operator, helicity is the generator of the electromagnetic duality transformation $D_\theta$, which acts on the fields as 
\begin{equation}\nonumber
\begin{split}
\EE\rightarrow \EE_\theta&=\EE\cos\theta  - Z\HH\sin\theta , \\
Z\HH\rightarrow Z\HH_\theta&=\EE\sin\theta + Z\HH\cos\theta.
\end{split}
\end{equation}

The relationship between helicity and duality is the same as, for example, angular momentum and rotations:
\begin{equation}
	R_z(\theta)=\exp(-i\theta J_z),\ D_\theta=\exp(-i\theta\Lambda).
\end{equation}

\begin{figure}[h!]
\includegraphics[width=\linewidth]{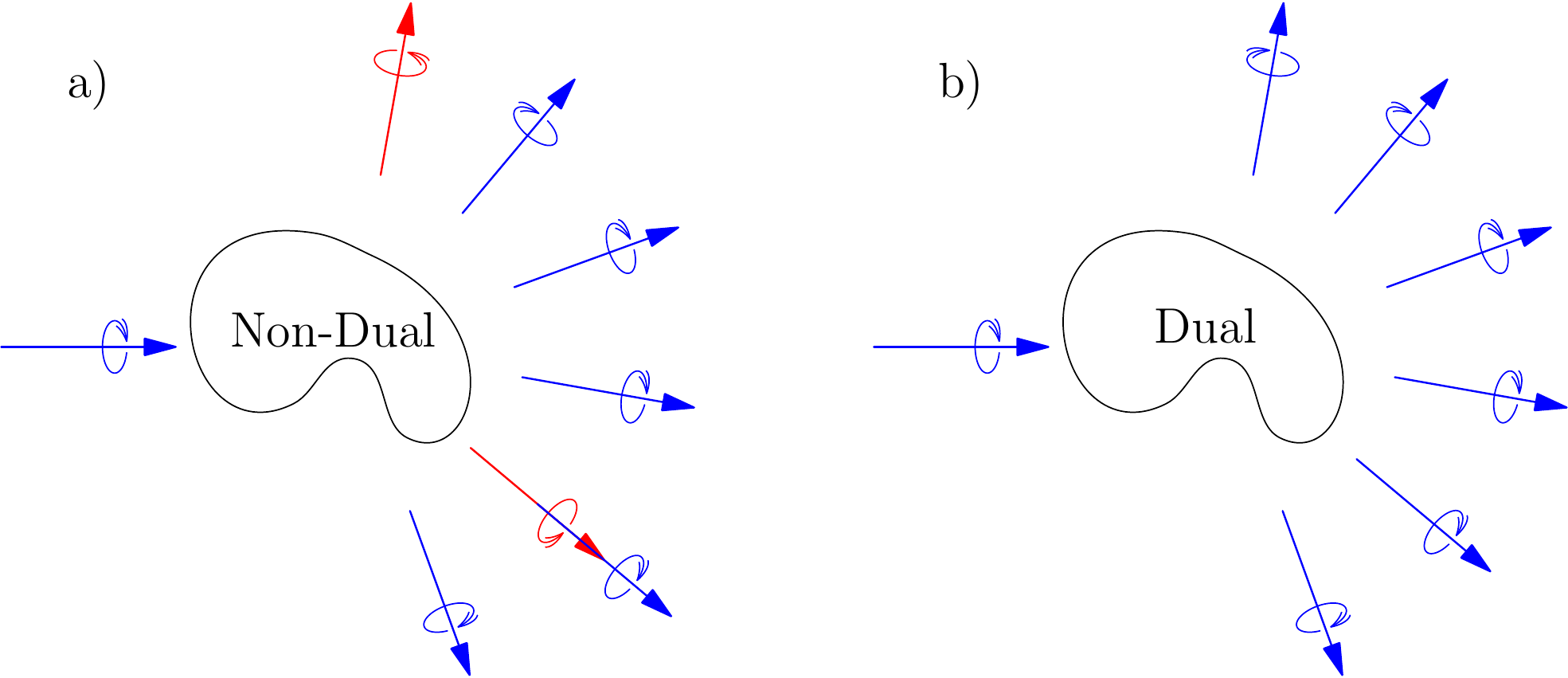}
\caption[Helicity preserving versus helicity non-preserving scatterers.]{\label{fig:dnd}(a): The helicity of an electromagnetic field is not preserved after interaction with a non-dual symmetric object. An incoming field with well defined helicity, in this case a single plane wave of definite polarization handedness (blue), produces a scattered field that contains components of the opposite helicity (red). The helicity of the scattered field in panel (a) is not well defined because it contains plane waves of different helicities. (b) Helicity preservation after interaction with a dual symmetric object. The helicity of the scattered field is well defined and equal to the helicity of the incident field.}
\end{figure}

The conditions for duality symmetry of a system described by the macroscopic equations can be found among other places in \onlinecite[Eq. 13]{FerCor2013}. The conditions for a dipolar scatterer to be dual symmetric are those in Eq. (\ref{eq:dual_dipolar}) of the main text. Figure \ref{fig:dnd}) illustrates scattering off dual and non-dual objects. The consideration of helicity and duality in light matter interactions facilitates their study by means of symmetries and conservations laws \cite{FerCorTHESIS}. 

\section{Helicity preservation in optical activity}\label{app:optact}

According to equations (\ref{eq:theta}) and (\ref{eq:beta}) of the main text, the polarization rotation angle can be computed as
\begin{equation}
\label{eq:betaapp}
\beta_\p^{\pbar}=\frac{1}{2}\arg{\left({\Fm}{\Fp}^*\right)}-\alpha_\p.
\end{equation}
The optical activity requirement of constant angle of rotation can be expressed as
\begin{equation}
	\frac{\partial \beta_\p^{\pbar}}{\partial \alpha_\p}=0,
\end{equation}
which implies through Eq. (\ref{eq:betaapp}) that
\begin{equation}
	\frac{\partial \arg{\left({\Fm}{\Fp}^*\right)}}{\partial \alpha_\p}=2.
\end{equation}
This means that $\left({\Fm}{\Fp}^*\right)$ must be of the form
\begin{equation}
		\label{eq:form}
		\begin{split}
				\left({\Fm}{\Fp}^*\right)&=\rho_\p^\pbar\exp(i(2\alpha_\p+\tau_\p^\pbar)),\\\textrm{ where }\frac{\partial \tau_\p^\pbar}{\partial \alpha_\p}&=0 \textrm{ and } \rho_\p^\pbar\textrm{ is real}.
		\end{split}
\end{equation}

Let us now expand the term $\left({\Fm}{\Fp}^*\right)$ using Eq. (\ref{eq:tx}) of the main text:
\begin{equation}
\label{eq:eler}
\begin{split}
	&\left({\Fm}{\Fp}^*\right)=\\
	&\ccc(\aaa)^*+\ddd(\aaa)^*\exp(i2\alpha_\p)+\ccc(\bbb)^*\exp(-i2\alpha_\p)+\ddd(\bbb)^*,
\end{split}
\end{equation}
and factor out the term $\exp(i2\alpha_\p)$
\begin{equation}
\label{eq:eler2}
\begin{split}
	\left({\Fm}{\Fp}^*\right)&=\\
	\exp(i2\alpha_\p)\big[&\ccc(\aaa)^*\exp(-i2\alpha_\p)+\\&\ddd(\aaa)^*+\\&\ccc(\bbb)^*\exp(-i4\alpha_\p)+\\&\ddd(\bbb)^*\exp(-i2\alpha_\p)\big].
\end{split}
\end{equation}

Since $\aaa$, $\bbb$, $\ccc$ and $\ddd$ are independent of $\alpha_\p$, the only term in Eq. (\ref{eq:eler2}) that has the form mandated by Eq. (\ref{eq:form}) is $\ddd(\aaa)^*\exp(i2\alpha_\p)$. In order to eliminate the other terms without eliminating the compliant one, we need to set $\bbb=\ccc=0$. We hence conclude that a polarization transformation consistent with optical activity is possible if and only if helicity is preserved.

When $\bbb=\ccc=0$ we have that
\begin{equation}
	\Fp=\aaa\exp (-i\alpha_\p),\ \Fm=\ddd\exp (i\alpha_\p),
\end{equation}
which, after substitution in Eqs. (\ref{eq:theta}), (\ref{eq:beta}) and (\ref{eq:eta2}) of the main text result in
\begin{equation}
	\beta_\p^{\pbar}=\frac{1}{2}(\arg\ddd-\arg\aaa), \ \eta_\p^\pbar=\frac{2|\ddd(\aaa)^*|}{|\ddd|^2+|\aaa|^2},
\end{equation}
which shows that $\eta_\p^\pbar$ is independent of $\alpha_\p$ as well.

\section{Ellipticity results}\label{app:eta}
\begin{figure}[ht!]
\includegraphics[width=0.8\linewidth]{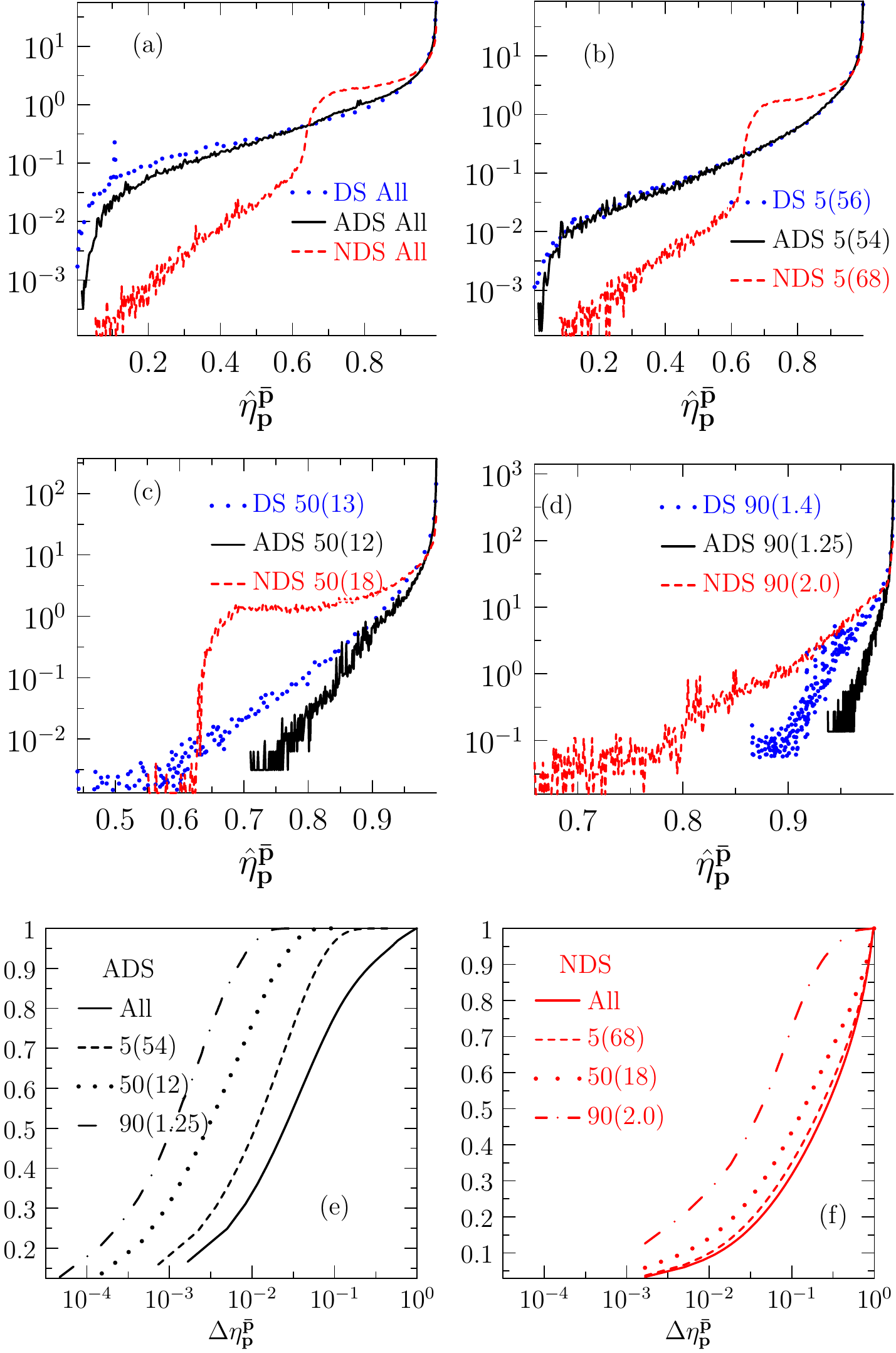}
\caption{\label{fig:etapic} Panels (a-d): Histograms of the average ellipticities $\etahat$. The different plots contain data for the different settings of the scattering threshold (see main text or the legend of Fig. \ref{fig:kpi}). Panels (e,f): cumulative histograms of $\Deltaeta$ for the ADS and NDS.}
\end{figure}

\end{document}